\documentclass[reprint,superscriptaddress,amssymb,amsmath,aps,showpacs,10pt,floatfix,pra,longbibliography]{revtex4-1}
\def\cm{cm$^{-1}$}
\usepackage{graphicx}%
\usepackage{color}
\usepackage{epstopdf}
\usepackage{amssymb}
\usepackage{amsmath}
\usepackage{amsfonts}
\usepackage{color}
\usepackage[colorlinks,bookmarks=false,citecolor=darkblue,linkcolor=red,urlcolor=blue]{hyperref} 
\definecolor{darkred}{rgb}{0.7,0.0,0.0}

\definecolor{darkblue}{rgb}{0,0.02,0.45}
\def\cdbl{\color{darkblue}}
\definecolor{darkgreen}{rgb}{0.02,0.45,0.0}

\definecolor{violet}{rgb}{0.8,0.2,0.6}

\newcommand{\av}{\mathbf a}
\newcommand{\bv}{\mathbf b}
\newcommand{\cv}{\mathbf c}

\begin{document}
\title{Optical signatures of phase transitions and structural modulation \\ in elemental tellurium under pressure}

\author{Diego Rodriguez}
\affiliation{1. Physikalisches Institut, Universit{\"a}t Stuttgart, 70569 Stuttgart, Germany}

\author{Alexander A. Tsirlin}
\email{altsirlin@gmail.com}
\affiliation{Experimental Physics VI, Center for Electronic Correlations and Magnetism, Augsburg University, 86159 Augsburg, Germany}

\author{Tobias Biesner}
\affiliation{1. Physikalisches Institut, Universit{\"a}t Stuttgart, 70569 Stuttgart, Germany}

\author{Teppei Ueno}
\affiliation{Research Institute for Interdisciplinary Science, Okayama University, Okayama, 700-8530, Japan}

\author{Takeshi Takahashi}
\affiliation{Research Institute for Interdisciplinary Science, Okayama University, Okayama, 700-8530, Japan}

\author{Kaya Kobayashi}
\affiliation{Research Institute for Interdisciplinary Science, Okayama University, Okayama, 700-8530, Japan}

\author{Martin Dressel}
\affiliation{1. Physikalisches Institut, Universit{\"a}t Stuttgart, 70569 Stuttgart, Germany}

\author{Ece Uykur}
\email{ece.uykur@pi1.physik.uni-stuttgart.de}
\affiliation{1. Physikalisches Institut, Universit{\"a}t Stuttgart, 70569 Stuttgart, Germany}

\date{\today}

\begin{abstract}

A room-temperature infrared spectroscopy study of elemental tellurium at pressures up to 8.44\,GPa in the energy range $0.015-2$\,eV is reported. Optical signatures of the high-pressure polymorphs are investigated and compared to the results of density-functional band-structure calculations. A Drude peak is first seen in the optical conductivity around 3.5\,GPa indicating a semiconductor-to-metal transition within trigonal Te-I. A sharp increase in the Drude spectral weight and dc-conductivity around 4.3\,GPa signals the transformation toward the triclinic Te-II polymorph. An absorption peak around 0.15\,eV appears above 5\,GPa concomitant with the gradual transformation of Te-II into the structurally similar but incommensurately modulated Te-III. Microscopically, this peak can only be reproduced within a sufficiently large commensurate approximant, suggesting the low-energy optical response as a fingerprint of the structural modulation.
\end{abstract}

\maketitle

\textit{\cdbl Introduction.}~Pressure-induced structural evolution of elemental solids is subject to long-standing investigations~\cite{Cannon1974, Pistorious1976} motivated by interesting electronic properties that can be reached in high-pressure polymorphs. Tellurium is one of the chemical elements that received renewed attention as a constituent of many topologically non-trivial materials~\cite{Chen2015, Tran2014, Xi2013, Wang2016, Jiang2017}. Theoretical studies suggested that even pure tellurium -- along with selenium, its neighbor in group VI of the periodic table -- can possess topological properties, but only under external pressure~\cite{agapito2013, Hirayama2015}. This triggered further high-pressure experiments~\cite{Nakayama2017}. 

External pressure can be used to fine tune band structures toward the desired topological regime, but it can also lead to abrupt changes in the crystal structure. These changes will sometimes result in symmetry lowering and increasing structural complexity~\cite{McMahon2006}, which is often intertwined with changes in physical properties, such as metallization and superconductivity~\cite{Shimizu1998, Eremets2001}. 

\begin{figure}
\centering
\includegraphics[width=1\linewidth]{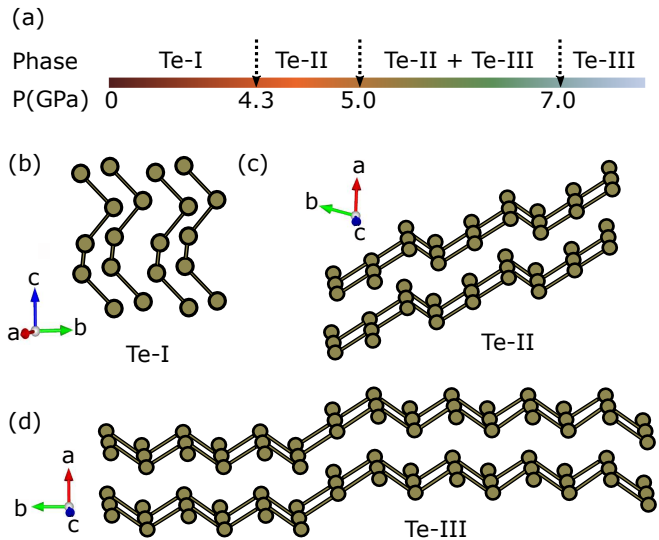}%
\caption{(a) Successive phase transitions of tellurium. Pressures are given as observed in the optical experiment also consistent with the presented XRD studies~\cite{hejny2003, hejny2004} (b) Te-I phase consist of the helical chains with only two covalent bonds formed by each tellurium atom. (c) Te-II phase organized as sheets with four-fold coordinated tellurium atoms with $2-1-2-1$ zigzag pattern. (d) Te-III phase built by the sheets with four-fold coordinated tellurium atoms with a complex pattern sequence with incommensurate periodicity.  }
\label{tellurium}%
\end{figure}

Elemental tellurium is a prominent example of this structural complexity in a chemically simple compound. Its high-pressure phases were controversially discussed in the literature, with the current consensus on the transformation of trigonal Te-I into triclinic Te-II around 4\,GPa~\cite{hejny2004} followed by a gradual transformation of Te-II into Te-III with the incommensurately modulated monoclinic crystal structure~\cite{hejny2003}. The Te-II $\rightarrow$ Te-III transformation starts around 4.5\,GPa and should be completed by 8\,GPa according to single-crystal x-ray diffraction data~\cite{hejny2004}. The Te-IV phase speculated in some of the earlier studies is concluded to be non-existent~\cite{Marini2012}, whereas the incommensurate Te-III structure is stable in a broad pressure range up to 30\,GPa before giving way to Te-V~\cite{hejny2004} (Fig.~\ref{tellurium}(a)). 

Although many inorganic and organic compounds are known to show incommensurately modulated crystal structures~\cite{Cummins1990, Lovelace2008}, simple elements joined the trend only more recently~\cite{Nelmes1999, Kenichi2003, Kume2005, Degtyareva2005, Fujihisa2007}. Interestingly, in chalcogens and halogens, which are non-metallic at ambient pressure, incommensurate structures will only form upon sufficient compression and \textit{after} metallic state is reached. This can be compared (and possibly contrasted) with the ambient-pressure behavior of metals that may reduce metallicity upon forming incommensurate charge-density waves. Experimental information on the electronic structure of the incommensurate high-pressure phases is scarce, though, owing to the limitations of experimental probes at such high pressures.

Elemental tellurium shows successive structural transitions and enters its incommensurate Te-III phase well below 10\,GPa, so we chose it for the high-pressure infrared study. Topological aspects of pressurized Te-I are covered in a separate publication that reports on signatures of the Weyl points above 3\,GPa~\cite{Rodriguez2019}. Here, we focus on the high-pressure polymorphs, trace the proposed structural transitions, and identify their unique fingerprints in the optical response.   

\textit{\cdbl Experimental details.}~Single crystals of tellurium were prepared by the Bridgman method explained elsewhere~\cite{Furukawa2017}. The crystals are easily cleaved at liquid nitrogen temperature perpendicular to the $b$-axis. High-pressure reflectivity measurements at room temperature were performed on $ac$-plane with the typical size of 160~$\mu$m$\times$170~$\mu$m$\times$60~$\mu$m. Measurements have been performed without utilizing a polarizer eflecting the average response over $ac$-plane. A screw-driven diamond anvil cell (DAC) equiped with the type-IIa diamond anvils with culet of 800~$\mu$m diameter is chosen for the measurements allowing us to reach pressures up to 8.44\,GPa. A CuBe gasket is preindented to $\sim$\,80~$\mu$m and a 250 $\mu$m diameter hole is drilled to be used as a sample chamber. Ruby spheres were loaded together with the sample as a pressure monometers, where the determination of the pressure inside the cell was achieved by monitoring the calibrated shift of the ruby R1 fluorescence line \cite{Mao1986}. 

The DAC was filled with finely ground CsI powder as a quasi-hydrostatic pressure-transmitting medium keeping a clean diamond-sample interface. The pressure gradient in the cell is monitored during the measurements via two different Ruby spheres sitting at the different side of the sample. At the low pressure regime (below 5-6\,GPa), no pressure gradient have been observed within the accuracy of the CCD spectrometer used in Ruby luminescence measurements (0.05\,nm / 0.15\,GPa). At the high pressure regime a gradient up to $\sim$0.7\,GPa is present. 

Room-temperature reflectivity measurements have been performed with a Hyperion infrared microscope coupled to a Bruker Vertex 80v Fourier transform infrared spectrometer. The pressure cell is attached to the microscope with a custom-made setup to control the position and the rotation of the pressure cell. Reflectivity spectra at the sample-diamond interface were collected between 100 and 20000 \cm\ and the CuBe gasket was used as a reference. Spectra have been corrected for the gasket reflection in the diamond anvil cell. The real part of the optical conductivity, $\sigma_1(\omega)$, was extracted via Kramers-Kronig (KK) analysis taking into account the sample-diamond interface \cite{Pashkin2006}. Simultaneous fits of the reflectivity and the optical conductivity by a Drude-Lorentz model ensure that the KK procedure has been carried out correctly. The spectra in the frequency range between 1700 and 2500 \cm\ are affected by the multiphonon absorptions of the diamond anvils, therefore, this energy range has been extrapolated linearly before the KK-analysis of the reflectivity. 

\textit{\cdbl Computational method.}~Relativistic band structures of Te-I, Te-II, and Te-III were calculated using the Wien2K code~\cite{wien2k} with the modified Becke-Johnson (mBJ) exchange-correlation functional~\cite{tran2009} for Te-I and generalized gradient approximation (GGA)~\cite{pbe96} for Te-II and Te-III. The use of mBJ, the functional optimized for semiconductors, was required, because relativistic GGA renders Te-I metallic already at ambient pressure. Experimental crystal structures were taken from Ref.~\onlinecite{keller1977} (Te-I) and Ref.~\onlinecite{hejny2004} (Te-II). In the case of Te-III, we adopted the lattice parameters from Table~\ref{tab:Te-III} and the modulation function given in Ref.~\onlinecite{hejny2003}. The modulation vector of Te-III is $(0,q,0)$ with $q=0.27\div 0.32$ depending on pressure~\cite{hejny2003}. Therefore, we used commensurate approximants with $q=\frac13\simeq 0.333$ ($t=0.1$) and $q=\frac27\simeq 0.286$ ($t=0.02$) in DFT calculations. The exact value of $t$ (initial phase of the modulation) has nearly no influence on the band structures and optical conductivity. However, we chose to avoid $t=0$, as it would lead to a centrosymmetric structure, whereas in general the Te-III structure violates inversion symmetry locally. 

Band dispersions were calculated on the well-converged $48\times 48\times 48$ $k$-mesh for Te-I and $24\times 6\times 24$ $k$-mesh for Te-II and Te-III. Denser meshes with $140\times 140\times 140$ points (Te-I) and $72\times 18\times 72$ points (Te-II, Te-III) were used for the calculation of the optical conductivity via the \texttt{optic} module~\cite{draxl2006} of Wien2K.


\begin{figure}
\centering
\includegraphics[width=1\linewidth]{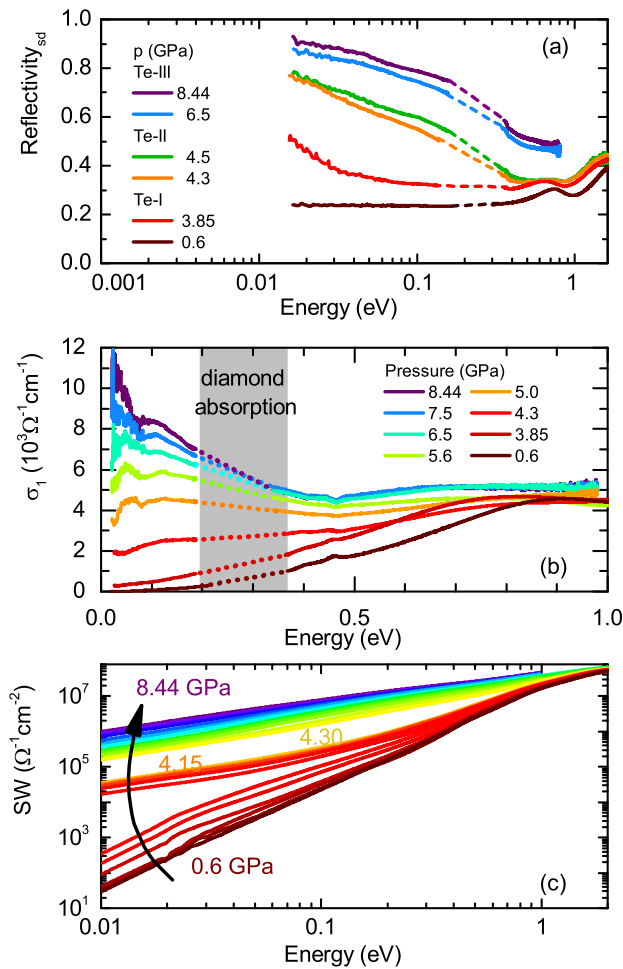}%
\caption{(a) Pressure-dependent reflectivity measured across the structural phase transitions. The shaded region is the diamond absorption range with the linear extrapolation of the reflectivity. (b) Pressure-dependent optical conductivity. (c) Spectral weight (SW) as a function of energy and pressure. While the Te-I phase can be identified with a slight red-shift of the high-energy bands in the optical conductivity and a small increase in the SW below 0.1~eV, the appearance of high-pressure polymorphs significantly modifies the optical conductivity, with a drastic increase of the spectral weight in a broader energy range. }
\label{pressure}%
\end{figure}

\textit{\cdbl Pressure-dependent optical spectra.}~In Fig.~\ref{pressure}(a), pressure-dependent reflectivity spectra are shown across the structural phase transitions. 
While an insulating behavior is observed at the lowest pressure, a small upturn indicating a Drude-like contribution can be seen at 3.85~GPa, still within the Te-I phase. With the appearance of the Te-II phase, the Drude-like contribution shows a sudden enhancement, whereas further increase in pressure results in a gradual increase of the metallicity. 

Similar behavior can be traced in the optical conductivity, Fig.~\ref{pressure}(b). At 0.6~GPa, it gradually decreases toward low energies and is eventually suppressed at a finite energy. This behavior reflects the presence of a band gap. On the other hand, at 3.85~GPa the optical conductivity extrapolating to zero frequency with a small increase of the low-energy spectral weight (SW) (Fig.~\ref{pressure}(c)), indicates the closing of the band gap and metallic nature of the system, which is also corroborated by the small upturn in the reflectivity at this pressure. These results indicate that at 3.85\,GPa tellurium is already metallic, but characterized by a low carrier density and a small Fermi surface that we ascribe to Te-I rather than Te-II according to the calculated band structures (see below).

Further compression leads to more drastic changes in the spectra. The sudden increase in the Drude-contribution to the reflectivity is paralleled by an abrupt increase in the low-energy optical conductivity. Above 5.0\,GPa, a peak-like structure evolves, and the dc-conductivity steadily increases. At even higher pressures, the increased Drude-like contribution screens out this low-energy absorption. The most abrupt change is observed between 4.15 and 4.3\,GPa, where the spectral weight, calculated as 
\begin{equation}
SW = \int_0^{\omega_c}\sigma_1(\omega)d\omega,
\label{SWeq}
\end{equation}
increases over a much larger energy range up to nearly 1\,eV, compared to the less prominent changes below 4.15\,GPa that are, moreover, restricted to lower energies. Therefore, our data indicate a transition around 4.3\,GPa with a pronounced increase in the carrier concentration and Fermi surface. Such a transition has to be accompanied by significant structural changes and corresponds to the transformation of Te-I into Te-II and Te-III, as we show below. 

\begin{figure}
\centering
\includegraphics[width=0.9\linewidth]{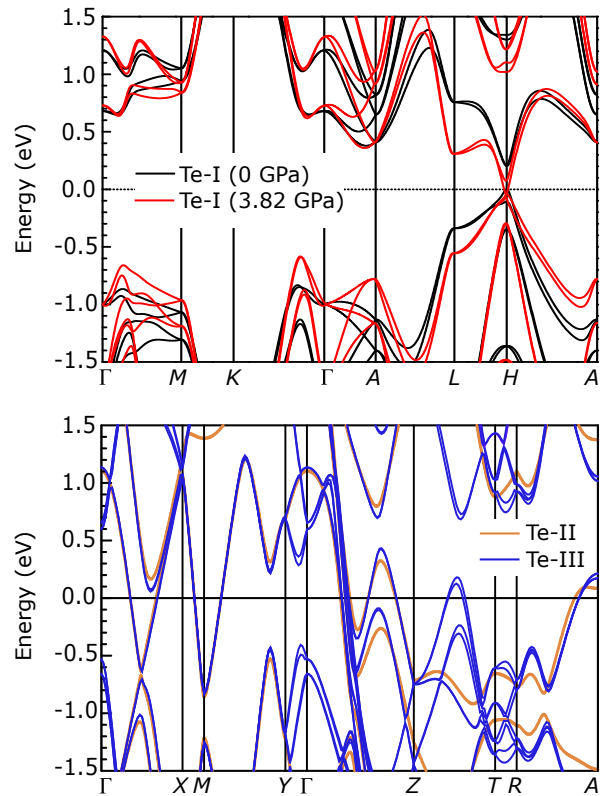}%
\caption{(a) Relativistic band structures of Te-I in the semiconducting (0\,GPa) and metallic (3.85\,GPa) states. Standard notation of $k$-points for trigonal symmetry is used. (b) Relativistic band structures of Te-II and Te-III ($q=\frac13$) with $k$-points given by $\Gamma(0,0,0)$, $X(0.5,0,0)$, $M(0.5,0.5,0)$, $Y(0,0.5,0)$, $Z(0,0,0.5)$, $T(0.5,0,0.5)$, $R(0.5,0.5,0.5)$, $A(0,0.5,0.5)$ in the reciprocal coordinates corresponding to the 5.06\,GPa lattice parameters of Te-III from Table~\ref{tab:Te-III} ($b$ is multiplied by 3 to account for $q=\frac13$). The 4.5\,GPa Te-II structure is transformed to the same setting using $\av=\av_t$, $\bv=(\av_t+\bv_t+\cv_t)/2$, and $\cv=\cv_t$, where $\av_t$, $\bv_t$, and $\cv_t$ are lattice vectors of the $I\bar 1$ cell given in Ref.~\onlinecite{hejny2004}.}
\label{DFT}%
\end{figure}

\begin{table}
\caption{\label{tab:Te-III}
Structural parameters of Te-III used for constructing the commensurate approximants. The data follow Ref.{~\cite{hejny2003}} with exact values provided by Prof. M.I. McMahon (private communication). The parameters of the modulation function were fixed to $B_{1x}=0.0315$ and $B_{1z}=0.1025$.
}
\begin{ruledtabular}
\begin{tabular}{ccccc}
 Pressure (GPa) & $a$ (\r A) & $b$ (\r A) & $c$ (\r A) & $\beta$ (deg) \\\hline
 5.06           & 4.0657     & 4.7376     & 3.1028     & 112.864       \\
 6.52           & 3.9908     & 4.7370     & 3.0818     & 113.332       \\
 8.49           & 3.9193     & 4.7334     & 3.0612     & 113.539       \\
\end{tabular}
\end{ruledtabular}
\end{table}

\textit{\cdbl Band structures.}~At ambient pressure, Te-I is a semiconductor (Fig.~\ref{DFT}) with the band gap of about 0.3\,eV at the $H$-point. This gap is closed under pressure leading to the band crossing and an associated small Fermi surface developing at 3.85\,GPa. In contrast, both Te-II and Te-III are robust metals with bands crossing the Fermi level in several parts of the Brillouin zone. Their metallicity is rooted in the drastic structural changes taking place upon the Te-I $\rightarrow$ Te-II transformation when the density increases by about 6\% with a concomitant increase in the coordination number of tellurium atoms. In contrast, the Te-II $\rightarrow$ Te-III transformation has no clear signatures in the pressure dependence of the unit cell volume~\cite{hejny2004}. 

The crystal structure of Te-I comprises helical chains with only two covalent bonds formed by each tellurium atom (Fig.\,\ref{tellurium}(b)), whereas Te-II and Te-III structures are built by sheets with four-fold coordinated tellurium atoms (Fig.\,\ref{tellurium}(c) and (d)). The only difference between Te-II and Te-III is in the organization of the layers that follow a regular $2-1-2-1$ zigzag pattern in the former, while showing a more complex sequence with incommensurate periodicity in the latter. The higher coordination number of tellurium renders both Te-II and Te-III robust metals with only a minor difference between their band structures (Fig.~\ref{DFT}(b)), which is mostly caused by the band folding in Te-III, owing to the twice larger number of atoms in the primitive cell (for $q=\frac13$).  

\begin{figure}
\centering
\includegraphics[width=1\linewidth]{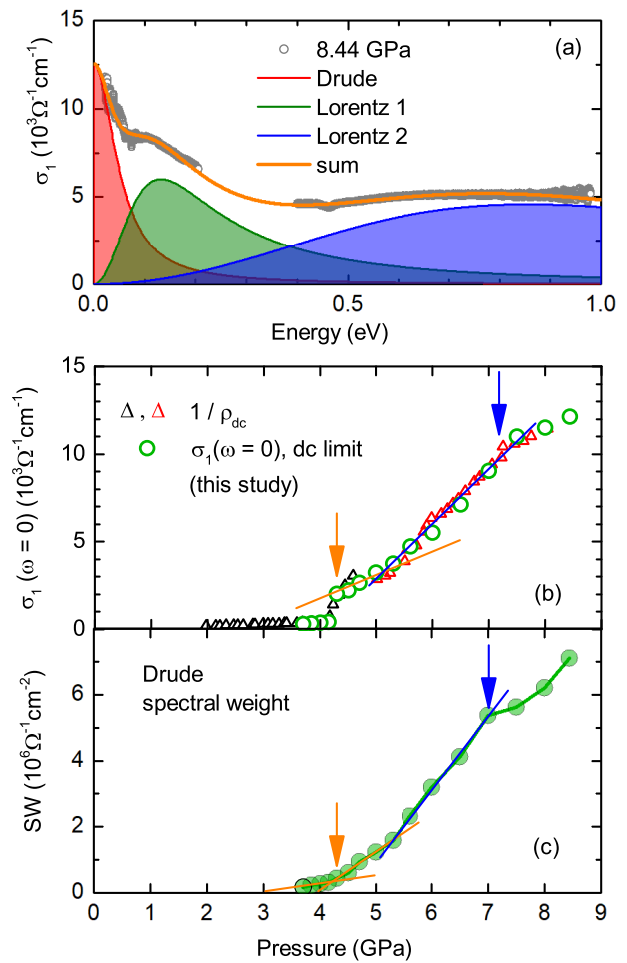}%
\caption{(a) Decomposition of the p = 8.44\,GPa optical conductivity into the Drude component for the free-carrier contribution and two Lorentz components for the MIR absorption (Lorentz 1) and high-energy absorption (Lorentz 2), respectively. (b) Optical conductivity in the $\omega\rightarrow0$ limit as a comparison to the dc resistivity measurements (black and red triangles at the low and high pressure range, respectively)~\cite{Ramasesha1991}, demonstrating the accuracy of the fit parameters. (c) Spectral weight of the Drude component. Drude SW is calculated from the individual Drude components obtained by fitting the optical conductivity, with the frequency limit of Eq.~\eqref{SWeq} taken as $\omega_c$ = 0.7\,eV that covers the entire Drude contribution. The orange arrows in (b) and (c) demonstrate the Te-I $\rightarrow$ Te-II transitions. The blue arrow shows the disappearance of the Te-II phase. The lines are guide to the eye.}
\label{SW}%
\end{figure} 

\textit{\cdbl Optical signatures of Te-III.}~Despite these similarities, we can clearly distinguish the formation of the incommensurate polymorph in the optical response of tellurium. To this end, we decompose the optical conductivity into the Drude component (free-carrier contribution) and two Lorentzian components due to the interband absorption, as shown in Fig.~\ref{SW}(a). In Fig.~\ref{SW}(b), optical conductivity in the dc-limit ($\omega=0$) is plotted along with the literature values obtained from the dc-resistivity. Their excellent match justifies our choice of the Drude contribution in the decomposition of $\sigma_1$. 

Already the Drude contribution tracks structural transformations of elemental tellurium. Below 4.3\,GPa, low dc-conductivity and small Drude SW indicate low carrier density in Te-I. A sudden jump in the dc-conductivity and the faster increase in the Drude SW, both shown with the orange arrows in Fig.~\ref{SW}, signal the transformation of Te-I into Te-II with its much larger Fermi surface and higher carrier concentration. Above 5.0\,GPa, the dc-conductivity increases further, whereas the Drude SW grows even faster up to about 7\,GPa, where a kink is observed (blue arrows in Fig.~\ref{SW}). We ascribe this regime to the gradual transformation of Te-II into Te-III.

\begin{figure}
\centering
\includegraphics[width=1\linewidth]{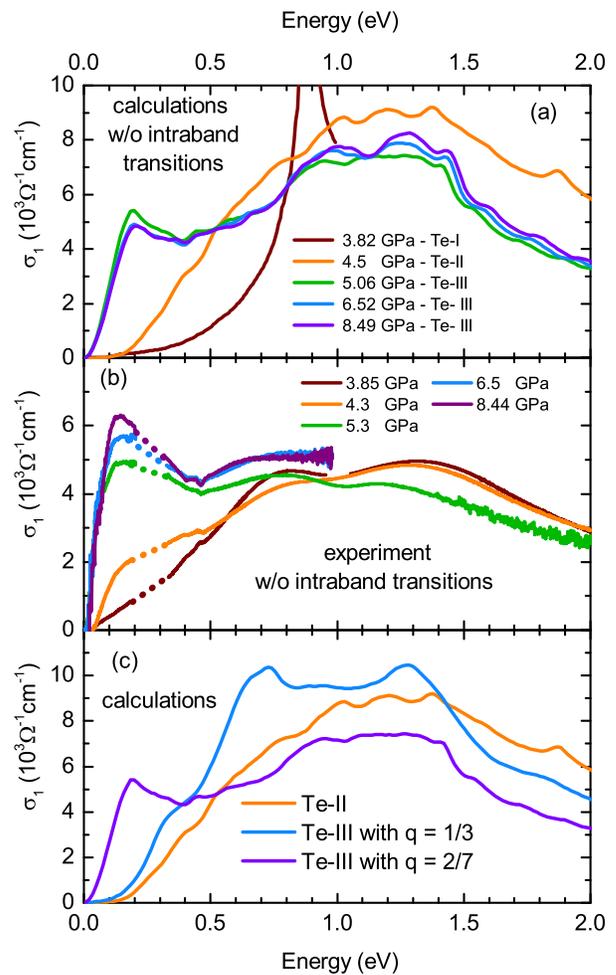}%
\caption{(a) Calculated optical conductivity for Te-I, Te-II, and Te-III ($q=\frac27$). (b) Experimental optical conductivity with the intraband contributions subtracted for a comparison to the calculations. (c) Calculated optical conductivity for Te-II and Te-III. In the Te-III case, two different modulation vectors ($q=\frac13$ and $q=\frac27$) have been used to demonstrate that the low-energy absorption band appears only with the increased periodicity of the structure. }
\label{calc}%
\end{figure}

Our assignment is supported by the evolution of the interband absorption shown in Fig.~\ref{calc}(b) after subtracting the Drude component. The interband absorption is represented by two Lorentzian contributions centered around 0.15\,eV and 1.0\,eV, respectively. The latter arises from multiple transitions between valence and conduction bands and appears in all three polymorphs of tellurium, while the former is rather unusual and indicates strong optical transitions at remarkably low energies. It becomes prominent above 5.0\,GPa only. Neither Te-I nor Te-II show this low-energy absorption in the calculated optical conductivity. In fact, even Te-III does not show a separate peak in this energy range when modeled within the $q=\frac13$ approximant (Fig.~\ref{calc}(c)), although its optical conductivity is somewhat higher than that of Te-II. On the other hand, the $q=\frac27$ approximant of Te-III does show a peak in the optical conductivity around 0.2\,eV, with the asymmetric shape that closely resembles our experimental data.

The origin of this low-energy peak is two-fold. First, the bands of Te-III are weakly split by the spin-orbit coupling and produce optical transitions that would not be possible in Te-II, where band splitting is forbidden by the inversion symmetry. Second, in Te-III several weakly spaced bands run parallel to each other along $\Gamma-Z$ (Fig.~\ref{DFT}(b)) and facilitate optical transitions at low energies. The number of these bands is the number of atoms per unit cell and increases upon going from $q=\frac13$ to $q=\frac27$, thus making the low-energy peak much more prominent in the latter case. Altogether, increased deviations from commensurability enhance the low-energy absorption peak. This peak serves as the most direct optical signature of the incommensurate Te-III phase.

\textit{\cdbl Discussion and summary.}~Optical data combined with the DFT calculations suggest that Te-I becomes metallic with a small Fermi surface and low carrier concentration before transforming into Te-II around 4.3\,GPa. Further compression leads to a gradual transformation of Te-II into Te-III, which is complete around 7\,GPa. At higher pressures, the dc-conductivity and Drude SW continue increasing, whereas the absorption peak around 0.2\,eV is nearly unchanged. This scenario is in excellent agreement with the XRD data that revealed first signatures of Te-II at 4.0\,GPa, first signatures of Te-III at 4.5\,GPa, and the complete transformation of Te-II into Te-III around 8\,GPa~\cite{hejny2003,hejny2004}. 

Despite their structural similarity, Te-II and Te-III exhibit a remarkably different optical response. Most notably, Te-III gives rise to a prominent low-energy absorption peak that we ascribe to the presence of an incommensurate structural modulation. In fact, aperiodic crystals have rarely been subject of optical studies, with the exception of quasicrystals~\cite{Homes1991, Burkov1992, Basov1994, Basov1994a, Timusk2013}, where effects on carrier density have been discussed and the weak Drude conductivity is concluded to be masked by the stronger interband absorption. Moreover, the weak localization effects in quasicrystals giving rise to decreasing low-energy optical conductivity has been pointed out. Aperiodic structures of simple elements may show interesting parallels to quasicrystals and await detailed investigation, but even from the data at hand we can conclude that the incommensurate Te-III polymorph shows increased metallicity and enhanced optical transitions at low energies compared to the structurally similar commensurate Te-II. This is at odds with ambient-pressure incommensurate charge-density-wave phases that show lower metallicity than their commensurate counterparts. 

In summary, we performed high-pressure infrared measurements on elemental tellurium over a broad frequency and pressure range. The structural phase transitions have been identified in the optical spectra, and the associated electronic transitions have been studied. While tellurium becomes metallic already in the Te-I phase, electronic structure and hence the optical properties change more abruptly with the appearance of Te-II. A sudden jump in the optical conductivity and the increase in the spectral weight over a much larger frequency range are the fingerprints of this phase. Although Te-III is structurally similar to Te-II, the presence of an incommensurate structural modulations results in a low-energy absorption peak that partially masks the Drude-like free-carrier contribution. At higher pressures, this absorption peak is screened out by the increased intraband contribution due to a higher carrier density. 

\begin{acknowledgments}
Authors acknowledge the fruitful discussions with Artem V. Pronin and Sascha Polatkan; technical support from Gabriele Untereiner. We are grateful to Malcolm McMahon for sharing his X-ray data for our calculations. Work in Okayama is supported by the Grant-in-Aid for Scientific Research (Grant Numbers: 18K03540, 19H01852). E.U. acknowledges the European Social Fund and the Baden-Württemberg Stiftung for the financial support of this research project by the Eliteprogramme.
\end{acknowledgments}

\bibliography{PRBTe}

\end{document}